\documentclass[aps,twocolumn,showpacs,floatfix]{revtex4}
\usepackage{graphicx,epsf,bm}

\let\oldmarginpar\marginpar
\renewcommand\marginpar[1]{\-\oldmarginpar[\raggedleft\small\sf #1]{\raggedright\small\sf #1}}

\newcommand{\FigWidth}{\columnwidth}

\def\S{{\tilde \Sigma}}

\begin{document}
\title{Ward identities for strongly coupled 
Eliashberg theories}
\author{Andrey V. Chubukov$^{1,2}$}
\affiliation{$^1$Department of Physics, University of Wisconsin,
Madison, WI 53706\\
$^2$ Department of Physics, University of Maryland, College Park, MD 20742-4111}\begin{abstract}
We discuss Ward identities for strongly interacting fermion systems described 
 by Eliashberg-type theories.
We show that Ward identities are not in conflict with Migdal theorem and
 derive  diagrammatically Ward identity for a charge 
 vertex both in a Fermi liquid, and  when a
 Fermi liquid is destroyed at quantum criticality. We argue that
  Ward identity for the spin vertex cannot be obtained within 
  Eliashberg theory.
\end{abstract}


\maketitle

\section{introduction}

Conservation laws set  powerful constraints on
 the forms of spin and charge response functions of interacting electron systems. The most useful constraints are associated with the conservation of the 
 the total charge and the total spin of fermions, $e_{tot}$ and 
$S^z_{tot}$. 
The time independence of the total charge and  spin 
 implies that the charge and spin susceptibilities, $\chi_c (q, \omega)$  and $\chi_s (q, \Omega)$, respectively, should vanish at 
 $q=0$ and any finite $\Omega$. This requirement imposes specific relations between self-energy and vertex corrections, known as Ward identities~
\cite{ward,ll,mahan}. 

The subject of this paper are  Ward identities for strongly interacting fermionic systems 
in which the self-energy is large, but is 
local -- it depends on frequency, but not on momentum, i.e.
 $\Sigma (k, \omega) \approx \Sigma (\omega)$. 
These theories have recently been introduced, both phenomenologically and  
 microscopically, in context of quantum criticality 
 in high $T_c$ and heavy fermion materials~\cite{NFL_at_QCP}.  
Examples include marginal Fermi liquid theory~\cite{mfl},
 gauge theories~\cite{ioffe}, and
critical theories of CDW~\cite{grilli} and SDW ~\cite{acs} 
transitions with the dynamical exponent $z >1$~~\cite{hertz}.
 In the last two cases, the locality of the problem
 is not imposed by the original Hamiltonian, but emerges
 at criticality due to the fact that for $z>1$,
collective spin or charge bosonic excitations
 are  slow modes compared to electrons.
The  locality of the self-energy 
 was first discovered by Eliashberg in the analysis of  
 electron-phonon problem~\cite{eliash},  
 and low-energy electron-boson 
theories with  $\Sigma (k, \omega) \approx \Sigma (\omega)$ bear his name.
The frequency-only dependent self-energy also emerges in 
 infinite $D$ theories of metal-insulator transition~\cite{kotliar} 
and in local theories of heavy-fermion quantum criticality~\cite{piers}.
However in these cases, the interaction is not small compared to a fermionic 
bandwidth, and  our analysis will not be applicable. 

The issue of Ward identities for Eliashberg theories is somewhat non-trivial as
these theories are justified by Migdal theorem that states that
vertex corrections $\delta \Gamma/\Gamma$ 
should generally be small compared to $\Sigma (\omega)/\omega$. Meanwhile, Ward identities imply that $\Delta \Gamma$ and $\Sigma (\omega)$ scale with each other. We show, however, that Migdal theorem and Ward identities 
just probe different limiting forms of the full, momentum and 
frequency dependent vertex $\Gamma (q, \Omega)$. Migdal theorem 
 states that for $v_F q >> \Omega$, $\delta \Gamma/\Gamma \ll \Sigma (\omega)/\omega$. Ward identity is applicable in the opposite limit of 
 $v_F q \ll \Omega$, and implies that in this limit, 
$\delta \Gamma/\Gamma \sim \Sigma (\omega)/\omega$.

We derive in this paper the  Ward identity for the charge vertex  near QCP.
We show that Ward identity holds both away from QCP, when the system
 is in the Fermi-liquid regime, and right at QCP, where the Fermi-liquid behavior may be destroyed by quantum fluctuations. We demonstrate that in the latter
 case, a derivation of Ward identity by explicit summation of diagrams is not straightforward and requires some care.
 
We also consider the 
  $SU(2)$ Ward identity for the spin vertex. We argue that 
 it holds near a CDW instability, but  
cannot be obtained within Eliashberg theory near a magnetic QCP. 
The argument is that the $SU(2)$ 
Ward identity is related to the conservation of 
 the total electron spin. Meanwhile, 
 Eliashberg theory near a SDW transition is based on the spin-fermion model
 in which  a soft collective bosonic mode in the spin channel is treated as 
a separate spin degree of freedom which can flip an electron spin. 
We argue that this effective two-fluid model is justified only 
 when the momenta of the collective bosonic mode is much larger than its frequency: $v_F q >> \Omega$. In this limit, Migdal theorem is  valid. 
 However, in the opposite limit  $v_F q << \Omega$, relevant to the $SU(2)$ 
Ward identity, the derivation of the two-fluid model near a magnetic QCP 
 breaks down, i.e., spin-fermion model is just inapplicable.
  
The structure of the paper is the following. 
We first discuss in the Sec.\ref{sec:2} how Ward identities are
 related to the 
conservation laws. In Sec \ref{sec:3} 
we demonstrate that  Ward identities are
 not in conflict with Migdal theorem. 
 In Sec. \ref{sec:4} we derive Ward identity for the charge vertex and
 discuss the difference between a Fermi liquid and a non-Fermi liquid. In Sec. \ref{sec:5} we discuss in some detail Ward identity for the spin vertex.

\section {Ward identities and the conservation laws}
\label{sec:2}

For $\Sigma (k, \omega) = \Sigma (\omega)$,  the full fermion-boson vertex $\Gamma$
 depends on incoming and outgoing fermionic frequencies $\omega$ and $\Omega + \omega$ ($\Omega$ is a bosonic frequency), and on a bosonic momentum $q$. Its dependence on fermionic momenta is small by Migdal condition and can be neglected.
The relevant Ward identity relates
the self-energy with  $\Gamma (q,\omega, \Omega)$
 at zero bosonic momentum $q$: $\Gamma (q =0,\omega, \Omega) = 
\Gamma (\omega, \Omega)$.
The identity implies that~\cite{ll,mahan} 
\begin{equation}
\Gamma (\omega, \Omega) = \frac{{\tilde \Sigma} (\omega + \Omega) - {\tilde \Sigma} (\omega)}{\Omega}
\label{1}
\end{equation}
where ${\tilde \Sigma} (\omega) = \omega + \Sigma (\omega)$
 We use the normalization in which 
 $\Gamma =1$ for non-interacting fermions. 
The role of  (\ref{1}) in enforcing the conservation laws becomes clear
when one considers a fully renormalized particle-hole polarization bubble
$\Pi (q, \Omega)$ at $q=0$. Both spin and charge 
susceptibilities scale with $\Pi (q, \Omega)$. In the RPA approximation, 
$\chi_s (q, \Omega) = 2 \Pi (q, \Omega)/(1 - U \Pi (q, \Omega))$, while
 $\chi_c (q, \Omega) = 2 \Pi (q, \Omega)/(1 + U \Pi (q, \Omega))$. Hence, 
if $\Pi (0, \Omega)$ vanishes, 
$\chi_s (0, \Omega)$ and $\chi_c (0, \Omega)$ also vanish.

\begin{figure}
\includegraphics[clip=true,width=\FigWidth,height=1.5in]{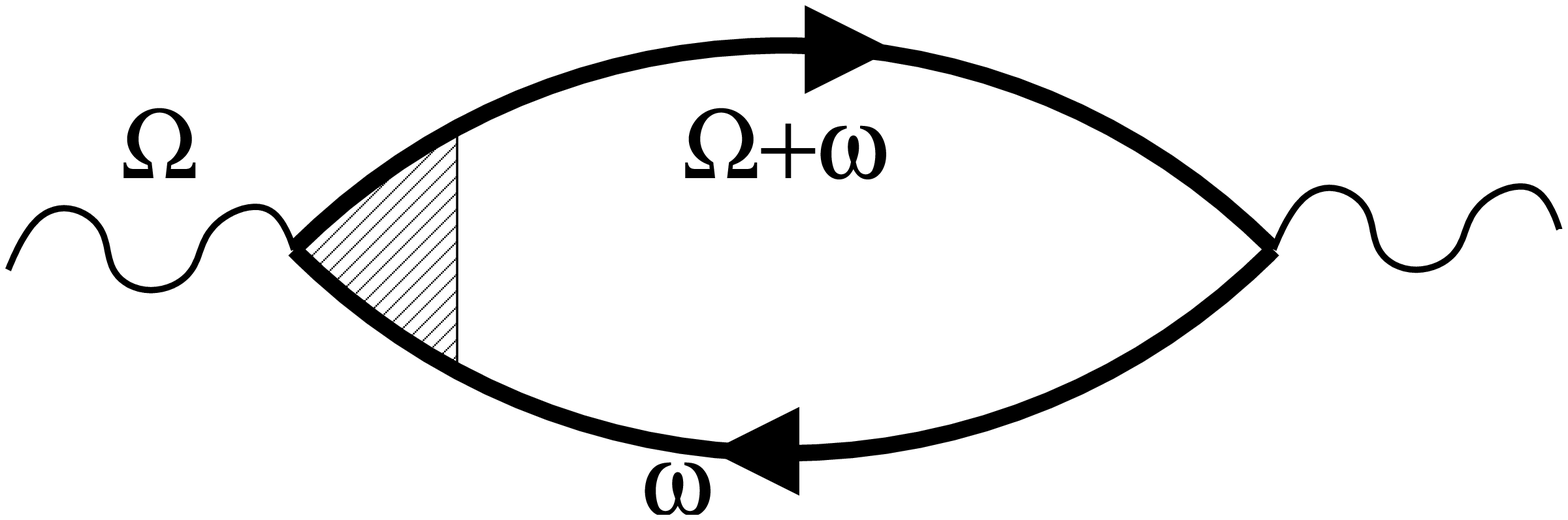}
\caption{The full particle-hole polarization bubble. 
 Solid lines -- full Green's functions, black triangle - the full vertex 
$\Gamma (\omega,\Omega)$.}
\label{fig1} 
\end{figure}

The proof that $\Pi (0, \Omega_m)$ vanishes for Eliashberg-type 
 theories has not been presented in the literature, so we discuss the 
computation in some detail.  The full polarization bubble 
 in Matsubara frequencies is given by 
the diagram in Fig.\ref{fig1} 
\begin{eqnarray}
\Pi (0, \Omega_m) &=& - \int \int \frac{d^D p d \omega_m}{(2\pi)^{D+1}} \times \nonumber \\
&& \Gamma (\omega_m, \Omega_m) G(p, \omega_m) G(p, \omega_m + \Omega_m)
\label{2}
\end{eqnarray}
where the fermionic Green's function $G(p, \omega_m)$ is given by
\begin{equation}
G (p, \omega_m) =  \left(i {\tilde \Sigma} (\omega_m) - \epsilon_p \right)^{-1}.\label{3}
\end{equation} 
Linearizing, as usual, the dispersion near the Fermi surface as
 $\epsilon_p = v_F (p-p_F)$, and replacing the momentum integral
$\int d^D p/(2\pi)^D$ by $ N_0 \int d \epsilon_p$,
 we obtain from (\ref{2}):
\begin{eqnarray} 
&&\Pi (0, \Omega_m) = - N_0  \int \frac{d \omega_m}{2\pi} \Gamma (\omega_m, \Omega_m) \times \nonumber \\
&&\int_{-W}^W d \epsilon_p \frac{1}{(\epsilon_p - i {\tilde \Sigma} (\omega_m)) (\epsilon_p - i {\tilde \Sigma} (\omega_m + \Omega_m))}
\label{4} 
\end{eqnarray}
where $W$ is of order of fermionic bandwidth.
It is common to set $W$ infinite in continuous theories.
We will see, however,  that it is important to keep $W$ finite
 at intermediate stages of calculations,
 and set $W = \infty$ only at the very end.
The need for this procedure follows from the fact 
 that at large frequencies 
${\tilde \Sigma} (\omega_m) \approx \omega_m$, $\Gamma \approx 1$, 
and the 2D integral over $d \epsilon_p d \omega_m$ formally diverges.
 A finite $W$ provides the physical 
regularization of the divergence as
 the linearization of the dispersion is only valid at $v_F (p -p_F) \leq E_F$.
 The frequency integral, on the other hand,  has to be evaluated in infinite limits. 
 
Integrating explicitly  over $\epsilon_p$ in (\ref{4}) we obtain
\begin{widetext}
\begin{equation}
\Pi (0, \Omega_m) = i N_0  \int^\infty_{-\infty} 
\frac{d \omega_m}{2\pi} \frac{\Gamma (\omega_m, \Omega_m)}{{\tilde \Sigma} (\omega_m + \Omega_m) - {\tilde \Sigma} (\omega_m)}~\log {\frac{(W - i\S (\omega_m + \Omega_m))(-W - i\S (\omega_m))}{(-W - i\S (\omega_m + \Omega_m))(W - i\S (\omega_m))}}
\label{5}
\end{equation}
\end{widetext}
One can easily make sure that  at $W \rightarrow \infty$, 
the frequency integral in (\ref{5}) 
is the sum of two terms: $\Pi (0, \Omega_m) =
\Pi_1 (0, \Omega_m) + \Pi_2 (0, \Omega_m)$. The first contribution
 comes from  {\it large} $\omega_m \gg \Omega_m$ for which 
$\S (\omega) \sim W$. Expanding
 in ${\tilde \Sigma} (\omega + \Omega) - {\tilde \Sigma} (\omega)$ in the logarithm, we obtain  
\begin{equation}
\Pi_1 (0, \Omega_m) =  4 N_0 \int_0^\infty \Gamma (\omega_m, \Omega_m) \frac{W}{W^2 + \S^2 (\omega_m)} \frac{d \omega}{2\pi}
\label{7}
\end{equation}
The integrand is $O(1/W)$ up to $\S (\omega) \sim W$, 
 and $O(W/\S^2 (\omega))$ at larger frequencies. This implies that 
 the integral is confined to $\omega$, for which $\S (\omega) \sim W$. 
 At these frequencies
 both the self-energy and the vertex  renormalization
 are irrelevant, i.e., $\Gamma (\omega_m, \Omega_m) =1$, and
 $\S (\omega_m) = \omega_m$. 
Performing the
 frequency integration in (\ref{7}) we then obtain 
\begin{equation}
\Pi_1 (0, \Omega_m) = 4 N_0 \int_0^\infty \frac{W}{W^2 + \omega^2_m}  \frac{d \omega}{2\pi} = N_0
\label{8}
\end{equation}
Another contribution comes from  
{\it small} frequencies $\omega_m \sim \Omega_m$, when $\omega_m$ and $\omega_m + \Omega_m$ have different signs, and the argument of the logarithm is 
$i \pi ~ (sgn (\omega + \Omega) - sgn \omega)$. The real part of the logarithm is $O(1/W)$ at small frequencies 
and can be neglected. Replacing the logarithm in (\ref{5}) by 
 its argument, we obtain
\begin{equation}
\Pi_2 (0, \Omega_m) = - N_0 \int_{- \Omega_m}^0 d \omega~\frac{\Gamma (\omega_m, \Omega_m)}{{\tilde \Sigma} (\omega_m + \Omega_m) - {\tilde \Sigma} (\omega_m)}~    
\label{8_1}
\end{equation}
For  definiteness, we set   $\Omega_m >0$. 
Using Ward identity, Eq. (\ref{1}), we immediately obtain
\begin{equation}
\Pi_2 (0, \Omega_m) = -N_0 \int_0^{\Omega_m} \frac{d \omega_m}{\Omega_m} =
 - N_0
\label{6}
\end{equation}
 Comparing (\ref{8}) and (\ref{6}), we see that 
the low frequency contribution, for which Ward identity is crucial, 
 cancels the one from high frequencies, and, as a result,  
$\Pi (0, \Omega_m)$ vanishes, as it indeed should.

\begin{figure}
\includegraphics[clip=true,width=\FigWidth,height=
1.2in]{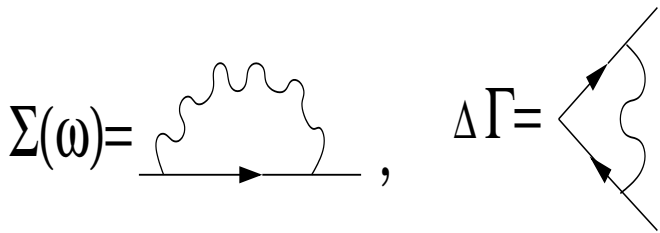}
\caption{The lowest-order diagrams for the fermionic self-energy and fermion-boson vertex. Wavy line is the bosonic propagator $\chi_l (\omega)$.}
\label{fig2} 
\end{figure}

\section{Diagrammatic derivation of Ward identity}
\label{sec:3}

\subsection{the relation to Migdal theorem} 

A more  subtle issue is how to prove Eq. (\ref{1}) diagrammatically.
A conventional recipe is to relate the equations 
 for the full vertex and  the fermionic self-energy~\cite{ll,mahan}.
At a first glance, this would contradict  
  Migdal theorem that states that in Eliashberg-type theories,
vertex corrections are much smaller than the self-energy.~\cite{migdal}.
We will demonstrate, however, that Migdal theorem   
is only valid for  $v_F q \gg \Omega$, 
while Eq. (\ref{1}) is valid for $v_F q \ll \Omega$. 
To understand the importance of the 
 $v_F q/\Omega$ ratio, note  that every 
 Eliashberg-type theory can be viewed as the interaction 
between fermions and the  
 ``local'' spin or charge  bosonic propagator  
$\chi_l (\Omega) = \int d q^{d-1} 
\chi (q, \Omega)$~\cite{ioffe,acs,cmhf,catherine,finn}.
The lowest order self-energy and  vertex corrections due 
to electron-boson interaction are shown in  Fig. \ref{fig2}.
For the self-energy, we have~\cite{acs}
\begin{equation}
\Sigma (\omega_m) \propto g^2 N_0  \int d \epsilon_k \int d \Omega_m 
 G_0(k, \omega^\prime_m + \Omega_m) 
\chi_l (\Omega_m)
\label{9_a}
\end{equation}
Evaluating the momentum integral using
\begin{equation}
\int d \epsilon_k  G_0(k, \omega_m) = \int \frac{d \epsilon_k}{i \omega_m - 
\epsilon_k} = -i \pi N_0 sgn \omega_m
\label{9b}
\end{equation}
we obtain
\begin{equation}
\Sigma (\omega_m) \propto g^2  \int_{-\omega_m}^0 d \Omega_m 
\chi_l (\Omega_m)
\label{9_c}
\end{equation}  
We see that the frequency  integral is confined to internal frequencies which are smaller than the external one. When $\omega_m$ is small, relevant $\Omega_m$ are also small, and away from the critical point 
$\chi_l (\Omega_m)$ can be approximated by $\chi_l (0)$. Then
\begin{equation}
\Sigma (\omega_m) = \lambda \omega_m, 
\label{9_d}
\end{equation}  
where $\lambda \propto g^2 N_0 \chi_l (0)$. When $\lambda \geq 1$,  $\partial 
\Sigma (\omega)/\partial \omega = \lambda$ is not small. 
Note that Eq. (\ref{9_c}) does not change if we re-evaluate the self-energy using the full Green's function $G(k, \omega) = (i {\tilde \Sigma} (\omega) - \epsilon_k)^{-1}$ for intermediate fermion. This follows from the observation that 
the relation (\ref{9b}) remains valid  for full $G$. 

For vertex correction, when both $q$ and $\Omega$ 
 are nonzero, we have 
\begin{eqnarray}
&&\delta \Gamma (q, \omega_m, \Omega_m) 
\propto i g^2 \int \int d^D k d \omega^\prime_m \times \nonumber \\
&&~G_0(k, \omega^\prime_m) G_0(k+q, \omega^\prime_m + \Omega_m) \chi_l (\omega_m - \omega^\prime_m)
\label{9}
\end{eqnarray}
This integral is ultraviolet convergent, and hence momentum integral can be extended to infinity. Integrating over momentum first, 
we obtain
\begin{equation}
\delta \Gamma (q, \omega_m, \Omega_m) 
\propto \frac{g^2 N_0}{{\sqrt{\Omega^2_m + (v_F q)^2}}}~\int_{-\Omega_m}^0 d \omega_m^\prime \chi_l (\omega_m - \omega^\prime_m)
\label{9a}
\end{equation}
We see that the integral over internal frequency $\omega^\prime_m$ is again 
confined
 to $|\omega^\prime_m| < \Omega_m$. Assuming that both $\Omega_m$ and $\omega_m$ are small, we again can approximate $\chi_l (\omega_m - \omega^\prime)$ by $\chi_l (0)$ and obtain
\begin{equation}
\delta \Gamma (q, \omega_m, \Omega_m)
 \propto \lambda \frac{\Omega_m}{\sqrt{\Omega^2_m + (v_F q)^2}}
\label{10}
\end{equation}
We  see that in the limit of vanishing $q$, $\Delta \Gamma \sim \lambda$ 
 i.e., $\delta \Gamma \sim \partial 
\Sigma (\omega)/\partial \omega$. This is consistent with Ward identity, Eq. (\ref{1}). 
 However, in the opposite limit $v_F q \gg \Omega$, $\delta \Gamma$ is reduced
 by $\Omega/v_F q$. For on-shell bosons, $\Omega = u q$, where
 $u$ is the effective ``sound'' velocity~\cite{comm}
  Hence,
 $\Delta \Gamma \sim \lambda u/v_F$. The ratio $u/v_F$ is a small parameter    
 in Eliashberg-type theories~\cite{mahan,migdal}, and
 the strong coupling limit of these theories 
 implies that $\lambda >1$, while $\lambda u/v_F \ll 1$. In this limit,
 the vertex for the scattering of fermions by on-shell bosons 
 is small, in agreement with Migdal theorem.

\begin{figure}
\includegraphics[clip=true,width=\FigWidth,height=1.2in]{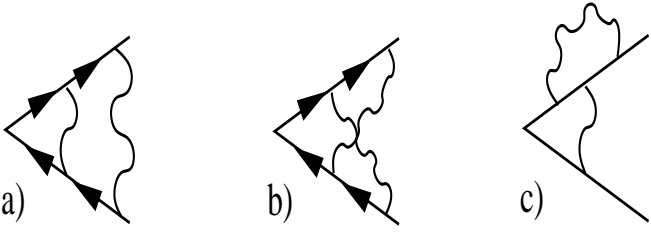}
\caption{ The second-order diagrams for the boson-fermion vertex
 $\Gamma (\omega, \Omega)$:  (a) ladder diagram, (b)  
crossed diagram, (b) the diagram with the renormalization of the side vertex.
 For Eliashberg theories, the ladder diagram is much larger than the 
other two.}
\label{fig3}
\end{figure}

\subsection{the selection of the diagrams}

The next issue is how to select  series of diagrams for 
 $\delta \Gamma (q, \omega_m,\Omega_m)$. A simple experimentation shows that 
 for $\lambda u/v_F \ll 1$, relevant diagrams for vertex renormalization
 form ladder series in which each 
 fermionic propagator is the full one.
 All non-ladder diagrams are small in $\lambda u/v_F$. 
To demonstrate this, compare the two second-order diagrams for 
  $\delta \Gamma (q, \omega_m, 
\Omega_m)$  - the ladder diagram and the crossed 
diagram (see  Fig.\ref{fig3} a, b). For both diagrams, we consider the limit
 when both $q$ and $\Omega_m$ are small, but $v_F q \ll \Omega_m$. 
 The ladder  diagram  $\delta \Gamma^{(2)}_{L}$ 
can be straightforwardly calculated in the same way as above and yields $\delta \Gamma^{(2)}_{L} \sim \lambda^2$. 
For the crossed diagram  $\delta \Gamma^{(2)}_{C}$, we have, setting 
 external $\omega_m =0$ and $|{\bf k}_{ext}| = k_F$, 
\begin{widetext}
\begin{equation}
\delta \Gamma^{{(2)}}_{C} \propto g^4 \int d^D k d^D p d \omega^\prime_m 
d \omega^{\prime \prime}_m G_0 ({\bf k}, \omega^\prime_m) G_0 ({\bf k} + {\bf q}, \omega^\prime_m + \Omega_m)~ G_0 ({\bf p}, \omega^{\prime \prime}_m)
 G_0 ({\bf k}_{ext} + {\bf k} - {\bf p}, \omega^\prime_m -\omega^{\prime \prime} _m)~\chi_l (\omega^\prime_m + \Omega - \omega^{\prime \prime}_m) ~
\chi_l (\omega^{\prime \prime}_m)
\label{11}
\end{equation}
\end{widetext} 
The integration over ${\bf k}$ and $\omega^\prime_m$ proceeds in the same way as before and yields a factor of $\lambda$. The integral comes from vanishingly small $\omega^\prime \leq \Omega_m$, hence $\omega^\prime$ can be neglected in the rest of the diagram. The remaining integral then reduces to 
\begin{eqnarray}
\delta \Gamma^{{(2)}}_{C} &\propto& \lambda \frac{g^2}{\chi_l (0)} \int d^D p d \theta 
  d \omega^{\prime \prime}_m \nonumber \\
&& \times   
G_0 ({\bf p}, \omega^{\prime \prime}_m)
 G_0 ({\bf l} - {\bf p},  -\omega^{\prime \prime} _m)~\chi^2_l (\omega^{\prime \prime}_m) 
\label{11_a}
\end{eqnarray} 
where ${\bf l} = {\bf k}_{ext} + {\bf k}$ is a vector whose length is $2k_F \cos \theta$.  
To avoid large denominators in the integrand in (\ref{11_a}),  
 both ${\bf p}$ and ${\bf l} - {\bf p}$ must be near the Fermi surface. 
For each given ${\bf l}$, this selects  special hot regions
 on the Fermi surface. The Fermi velocities at  ${\bf p}$ and ${\bf l} - 
{\bf p}$ in these two regions  are generically not aligned, 
 hence, in evaluating the momentum integral in 
(\ref{11_a}), one has to integrate independently over 
$\epsilon_{\bf p}$ and $\epsilon_{{\bf l} - {\bf p}}$. Using
 $\int d^D p \propto  (N_0/W) \int \int d 
\epsilon_{\bf p} d \epsilon_{{\bf l} -{\bf p}}$ and  
integrating each of the two Green's functions in (\ref{11_a}) 
 over its $\epsilon$ we obtain 
\begin{equation}
\delta \Gamma^{{(2)}}_{cr} \propto \lambda^2~ \chi^2_l (0) 
\int \frac{d \omega^{\prime \prime}_m}{W}~\left(f (\omega^{\prime \prime}_m)\right)^2
\label{11_b}
\end{equation} 
where we introduced 
$ f(\omega) = \chi_l (\omega)/\chi_l (0)$, $f(0) =1$.  

The  frequency integral converges at $\omega^{\prime \prime} \sim u p_F$
 such that the frequency integral gives
$u p_F/W \sim u/v_F$.  Substituting this into (\ref{11_b}), 
 we find that  $\delta \Gamma^{{(2)}}_{cr} \sim \lambda [\lambda  (u/v_F)]$, 
i.e., it  is small compared to ladder diagrams.
 One can verify
 that the same  smallness in $\lambda u/v_F$ emerges when one includes
 the renormalization of the side vertices in the ladder series (see Fig.\ref{fig3} c). 

\begin{figure}
\includegraphics[clip=true,width=\FigWidth,height=1.3in]{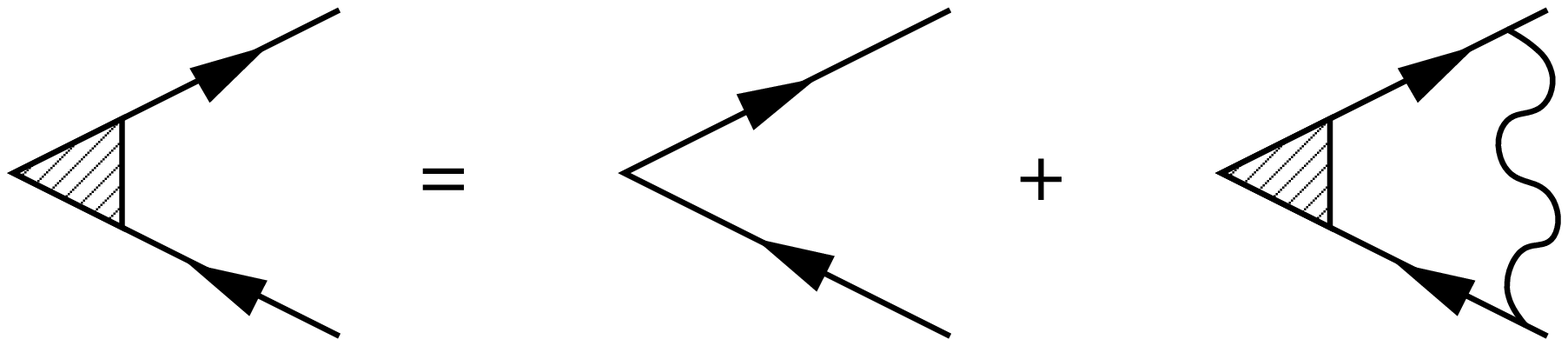}
\caption{The series of ladder diagrams for the full fermion-boson
 vertex. The diagrammatic series reduce to the
 integral equation for $\Gamma$.}
\label{fig4}
\end{figure}

To prove Eq. (\ref{1}) we now need to go beyond estimates and 
 explicitly sum up ladder series of diagrams.  
At this stage, we consider charge and spin vertices separately. 

\section{The charge vertex}
\label{sec:4}
 
\subsection{general proof}

We begin with the charge vertex $\Gamma_c (q=0)$. The bosonic $\chi_l (\omega)$ may  originate either in the charge or in the spin channel. In both cases, we
  define $\lambda$ via
$\Sigma (\omega_m) = \lambda \omega$ at small $\omega$. 
For charge fluctuations, $\lambda = 
g^2 N_0\chi_l (0)$, for spin fluctuations, $\lambda = 3 g^2 N_0 \chi_l (0)$.
Evaluating the lowest order  perturbative correction to the charge vertex, 
 we find that at small $\Omega_m$ and $\omega_m$, 
$\Gamma_c (\omega_m, \Omega_m) = 1 + \lambda$, i.e., vertex renormalization  
 contains {\it exactly} the same $\lambda$ independently on whether $\chi_l (\omega)$ is of spin or charge origin. 

Further, the ladder series of  of vertex correction 
diagrams  reduce to the integral equation 
 for $\Gamma_c (\omega_m, \Omega_m)$ in the form
\begin{eqnarray}
&&\Gamma_c (\omega_m, \Omega_m) = 1 +  i \lambda \int \int \frac{d \epsilon_k d\omega^\prime_m}{2\pi}~f(\omega_m -\omega^\prime_m) \times  \nonumber \\
&&\frac{\Gamma_c (\omega^\prime_m, \Omega_m)}{(i \S (\omega^\prime_m) - \epsilon_k)(i \S (\Omega_m + \omega^\prime_m) - \epsilon_k)} 
\label{14}
\end{eqnarray}
We show this graphically in Fig. \ref{fig4}. 

A similar equation can be obtained for the self-energy. 
Explicitly computing the prefactor in (\ref{9a}) and using the fact that the self-energy given by (\ref{9a}) does not change if we replace $G_0$ by full $G$, as long as $\Sigma$ depends only on frequency, we obtain 
\begin{eqnarray}
&&\S (\omega_m + \Omega_m) - \S (\omega_m) = \Omega_m  +
 i \lambda \int \int \frac{d \epsilon_k d\omega^\prime_m}{2\pi}~\times \nonumber \\
&&\frac{\S(\Omega_m + \omega^\prime_m ) - \S (\omega^\prime_m)}{(i 
\S (\omega^\prime_m) - \epsilon_k)(i \S (\Omega_m + \omega^\prime_m) - \epsilon_k)}~ f(\omega_m - \omega^\prime_m)
\label{15}
\end{eqnarray} 
Comparing Eqs. (\ref{15}) and (\ref{14}), we see that they are 
equivalent if
$\Gamma_c (\omega_m, \Omega_m) = (\S (\omega_m + \Omega_m) - \S (\omega_m))/\Omega_m$, i.e., when $\Sigma (\omega_m)$ and $\Gamma_c (\omega_m, \Omega_m)$ 
 are related by Ward identity -- Eq. (\ref{1}).

\subsection{perturbation theory order by order}
 
It is also 
instructive to analyze how Eq. (\ref{1}) is recovered by summing up
 vertex correction diagrams explicitly, order by order, and comparing the result with the self-energy. 
Order by order consideration was used to prove Ward 
identity for electron-phonon systems~\cite{mahan}. There, series of vertex corrections are geometrical and one can straightforwardly sum them up.
We will see that in our case, the series of vertex corrections are 
 geometrical   in the Fermi liquid regime, but become non-geometrical
 at the  QCP, where $\lambda = \infty$. In the last case, explicit order by order summation of perturbation series is not straightforward and requires some efforts.

\subsubsection{Fermi liquid}

In a Fermi liquid,
 $f(\omega) \approx 1$, and 
 $\S (\omega) = \omega_m (1 + \lambda)$. Then the leading vertex correction calculated with the full fermionic propagators is $\lambda/(1 + \lambda)$. 
One can easily make sure that 
 higher-order vertex corrections form simple 
geometrical series in $\lambda/(1+ \lambda)$. Adding them up, we immediately 
 obtain
\begin{eqnarray}
&&\Gamma_{c} (q=0)  = \frac{\lambda}{1+ \lambda} + \left(\frac{\lambda}{1+ \lambda}\right)^2 + ...= \frac{1}{1 - \frac{\lambda}{1+ \lambda}} \nonumber \\
&&= 1 + \lambda =
\frac{{\tilde \Sigma} (\omega + \Omega) - {\tilde \Sigma} (\omega)}{\Omega}
\label{13}
\end{eqnarray}

\subsubsection{quantum-critical point}

Suppose now that the system approaches a 
QCP at which the bosonic mode becomes massless. 
If the spatial dimension is below the critical one $d_{cr}$, 
 the local susceptibility $\chi_l (0)$ diverges at QCP what implies that 
 $\lambda -> \infty$ ($d_{cr}=3$ for Ornstein-Zernike form of $\chi (q, 0)$~\cite{acs,finn}). 
Suppose further that at QCP,
 the local susceptibility behaves as 
$\chi_l (\omega_m) \propto \omega_m^{-\gamma}$, and $\gamma >0$. 
Examples include 3D antiferromagnetic and ferromagnetic QCP~\cite{millis_roussev,bedell} and  marginal Fermi liquid theory~\cite{mfl}, where
 $\gamma = 0+$ (i.e., $\chi_l (\omega_m) \propto \log \omega$),
2D antiferromagnetic QCP where $\gamma =1/2$~\cite{acs,finn}, 
ferromagnetic QCP~\cite{cmhf,catherine,lonz,gorkov}
 and gauge theory~\cite{ioffe},
 where $\gamma = 1/3$,
 and corresponding CDW instabilities with the same $\gamma$~\cite{grilli,lonz}.  In all density-wave cases, upper critical dimension is $D_{cr} =3$. 

For $\chi_l (\omega_m) \propto \omega_m^{-\gamma}$, self-energy 
acquires a non-Fermi liquid form 
\begin{equation}
\Sigma (\omega_m) = \omega^{1-\gamma}_m \omega^\gamma_0,
\label{16}
\end{equation}
 where $\omega_0$ is the normalization factor. 
Ward identity implies that the fully renormalized vertex should behave as
 $\Gamma \propto \omega^{-\gamma}$, i.e., it should diverge at $\omega =0$.

Substituting this self-energy into (\ref{14})  and evaluating 
 vertex corrections iteratively, order by order,  we obtain for $\Omega_m, \omega_m >0$, 
\begin{widetext}
\begin{equation}
 \Gamma_c (\omega_m, \Omega_m)= 1 + (1-\gamma) \int_0^1 
\frac{d x}{(x +\frac{\omega_m}{\Omega_m})^\gamma}~\frac{1}{S(x)}\left[1 + 
(1-\gamma) \int_0^1 \frac{d y}{|x -y|^\gamma}~\frac{1}{S(y)}\left[1 + (1-\gamma) \int_0^1 \frac{d z}{|y -z|^\gamma}~\frac{1}{S(z)} +....\right]\right]
\label{18}
\end{equation}
\end{widetext}
where $S(x) = (\Omega_m/\omega_0)^\gamma + x^{1-\gamma} + (1-x)^{1-\gamma}$.
Evaluating the first few terms, we obtain for, e.g., $\gamma = 1/2$ and 
$\omega_m =\Omega_m =0$:  
\begin{equation}
 \Gamma_c (0, 0)= 1 + 0.7854 + 0.7578 + ...
\label{19}
\end{equation}
Clearly, the series are not geometrical, and first few terms 
do not give a hint what the full $\Gamma_c (0, 0)$ actually diverges.
This implies that  the direct, order by order summation of vertex correction diagrams is useless at the QCP.

Despite that  order by order summation fails,  infinite series in (\ref{18}) 
can indeed be evaluated explicitly. For this  we introduce $\phi (x)$ via
\begin{equation}
 \Gamma_c (\omega_m, \Omega_m) = 1 + (1-\gamma)\int_0^1 
\frac{d x}{(x +\frac{\omega_m}{\Omega_m})^\gamma} \frac{\phi (x)}{S(x)}
\label{20}
\end{equation}
 and observe that  $\phi (x)$ obeys the integral equation
\begin{equation}
\phi (x) = 1 + (1-\gamma) \int_0^1 \frac{d y}{|x -y|^\gamma}~
\frac{\phi (y)}{S(y)}
\label{21}
\end{equation}
One can verify that the solution of (\ref{21}) is 
$\phi (x) = S(x) (\omega_0/\Omega_m)^\gamma$. Substituting this $\phi (x)$ into
(\ref{20}) and performing the integration, we obtain 
\begin{equation}
\Gamma_c (\omega_m, \Omega_m)= 1 + \left(\frac{\omega_0}{\Omega_m}\right)^\gamma \left[\left(1 + \frac{\omega_m}{\Omega_m}\right)^{1-\gamma} - \left(\frac{\omega_m}{\Omega_m}\right)^{1-\gamma}\right]
\label{17}
\end{equation}
This is precisely the same result as one would obtain by 
 substituting the self-energy, Eq. (\ref{16}), into Eq. (\ref{1}).    
In particular,  when $\Omega_m$ tends to zero, and 
$\omega_m \sim \Omega_m$, the full charge vertex diverges as 
$\Omega_m^{-\gamma}$. 

\section{The spin vertex}
\label{sec:5}

Finally, we discuss the spin vertex $\Gamma_s$. Here the situation is more involved. 
When $\chi_l$ originates in the charge channel, as e.g., near CDW instability~\cite{grilli},
the spin factors in the self-energy and the vertex correction diagrams 
are the same (both equal to one),
 and Ward identity for $\Gamma_s$ 
readily follows from the summation of the ladder series of diagrams.
 However, if $\chi_l$ originates in the spin channel, as near a
 SDW instability~\cite{acs}, the interplay between the self-energy and
 the vertex corrections depends on the spatial anisotropy of $\chi_l$.  
If the SDW transition is of Ising type, i.e., only 
the $z-$ component of  $\chi_l$ is relevant,
 the spin summation yields the same factor $1$ both  for the
 self-energy and the vertex~\cite{millis_roussev,bedell,cmhf}, and the 
 ladder summation recovers Eq. \ref{1}. However, if the transition falls 
into Heisenberg universality class, the  summation over spin 
components in the first self-energy diagram 
yields a factor $\sum_\beta {\vec \sigma}_{\alpha,\beta} 
{\vec \sigma}_{\beta,\alpha} =3$, while the summation over spin components in 
the first vertex correction diagram yields a factor $(-1)$ as 
$\sum_{\alpha,\beta} \sigma^{i}_{\alpha,\beta} {\vec \sigma}_{\beta, \gamma} 
{\vec \sigma}_{\delta, \alpha} = - \sigma^{i}_{\delta, \gamma}$~\cite{cmhf}.
 As a result, the analogy between self-energy and vertex  correction 
diagrams is lost for $\Gamma_s$.
 In a Fermi liquid, ladder diagrams for $\Gamma_s$ now 
form series in $- (1/3) \lambda/(1+ \lambda)$, and the
 summation of the ladder series yields 
$\Gamma_s = 3 (1 + \lambda)/(3 + 4 \lambda)$ what is different from $1+ \lambda$ expected from Eq. \ref{1}. 

 The fact that  Eq. (\ref{1}) for $\Gamma_s$ is not recovered in the 
summation of the ladder diagrams 
 does not imply that the  Ward identity based on the
 conservation of the total $S^z$  is invalid. 
Rather, the non-equivalence between self-energy and vertex corrections is the
 consequence of the fact  that for spin vertex, 
   Eliashberg theory near the SDW instability
 is only applicable
 for $v_F q \gg \Omega$, but not at $\Omega \gg v_F q$ where Ward identity must be valid. Indeed,   
 the point of departure for the SDW Eliashberg  theory  is 
 the spin-fermion model~\cite{hertz,acs}. This model  assumes  
 that the low-energy 
physics is governed by the vector spin-spin 
interaction 
\begin{equation}
{\cal H}_{int} = g \sum_q {\bf s}(q) {\bf S} (q)
\label{22}
\end{equation}
   between  a fermion with a spin ${\bf s}(q) =
\sum_k c^\dagger_{k,\alpha} {\bf \sigma}_{\alpha, \beta} c_{k+q,\beta}$  
and a spin collective mode described by a bosonic field ${\bf S}(q)$.
 In other words, a static collective mode is formally treated as 
a separate spin degree of freedom. 
Within this description, the total spin of the system is
${\bf S}_{tot} (q) = {\bf s} (q) + {\bf S} (q)$. This total spin is indeed 
 a conserved quantity. However, the 
Ward identity, Eq. (\ref{1}), involves only the electron spin 
 as is evident from its relation to the electron particle-hole 
polarization bubble $\Pi (q, \Omega)$. If the interaction with 
the collective mode is
 of Ising type, there are no spin-flip processes
 between  $s^z (q)$ and $S^z (q)$, and 
 $s^z (q=0)$ and $S^z (q=0)$ are two independent  conserved quantities. 
  In this situation, Eq. (\ref{1}) is satisfied,
 as we have demonstrated above. However, for Heisenberg-type exchange 
with the collective mode, only ${\bf S}^z_{tot} (0)$ is conserved, but not 
 $s^z (0)$ as the electron spin can flip under the action of $S^+ (q) s^{-} (q)$ and transfer its $z$ component to 
${\bf S} (q)$.  As a result, within the spin-fermion model, 
 {\it there is no conservation of the electron spin}.

In reality, ${\bf S}$ is indeed an auxiliary field. Collective modes 
 are made out of electrons, so there is no independent 
spin degrees of freedom other than electron spins. 
The physics behind the effective  spin-fermion model with two 
spin degrees of freedom 
is the separation of scales. In perturbation series, 
the static part of the spin susceptibility
comes predominantly from electrons with energies comparable to the 
fermionic bandwidth $W \sim v_F p_F$, while the dynamics of the spin 
response function is dominated by the Landau damping and 
comes from fermions, with energies 
 comparable to the frequency of the collective mode  $\Omega$.
Eq. (\ref{22}) is valid if relevant $\Omega$ are much smaller than $W$. 
The largest bosonic momenta are of order $p_F$, hence the largest $\Omega$
 for typical, mass shell bosons 
 is of order $u p_F$. The condition $\Omega << W$ then implies $u << v_F$,
 which is  precisely the applicability condition for Eliashberg theory.
This condition, however, also implies that at
 small $q$, typical $\Omega \sim u q \ll v_F q$, i.e., 
 the limit
 of $q=0$ and finite $\Omega$,  relevant for the conservation laws, cannot
 be reached. 

\section{conclusions}
\label{sec:6}

To summarize, in this paper we considered Ward identities for 
 charge and spin vertices in strongly coupled Eliashberg-type theories
 in which $\Sigma (k, \omega) \approx \Sigma (\omega)$. We
  explicitly demonstrated how  Ward identities impose 
conservation laws and argued that Ward identities
 are not in conflict with Migdal theorem.
 For the charge vertex, we demonstrated that Ward identity 
is reproduced  by summing up ladder series of diagrams, both in the
 Fermi liquid regime, and when the Fermi liquid is destroyed at either CDW or SDW QCP.
 We found, however, that order-by-order summation is only useful in the 
Fermi liquid regime, but 
 becomes meaningless at the QCP. For the spin vertex, we 
 found that Ward identity 
 is reproduced near a CDW transition and near an Ising-type SDW transition, 
but cannot be reproduced  near a Heisenberg-type SDW transition. 
 We argued that this is the consequence of the fact that 
 Eliashberg theory near a Heisenberg SDW transition is only valid
 when the bosonic momentum is much larger than the bosonic frequency, 
$v_F q > \Omega$. Ward identity, on the contrary, 
is valid in the the opposite limit 
$v_F q < \Omega$, where Eliashberg theory is inapplicable.

This work was supported by NSF DMR
0240238. I thank  Ar. Abanov, C. Castellani, C. DiCastro, M. Grilli, 
D. Maslov,  C. Pepin, J. Rech and R. Ramazashvili  for useful conversations.


\begin{thebibliography}{20}
\bibitem{comm} For overdamped, diffusive bosons
 $u = u_q$ by itself scales as some power of $q$ (as $(q-q_c)^2$ near 
 a critical point at a finite $q_c$ and as $q^3$ near a critical point 
 at $q=0$). In this situation, the choice of 
 the effective velocity $u$ depends on what kind of problem is considered.
Thus, for a pairing problem near QCP, relevant frequencies $\Omega \sim u_q q$
 are of order of
 fermion-boson interaction $g$~\protect\cite{ioffe,grilli,acs,cmhf,catherine,finn,lonz}), hence
 relevant momenta $q_{typ}$ are determined from $u_{q_{typ}} q_{typ} \sim g$.  
In all cases studied so 
far~\protect\cite{ioffe,grilli,acs,cmhf,catherine,finn,lonz,gorkov})
 Eliashberg condition  $\lambda u/v_F \ll 1$ is valid when $g \ll W$.
\end{thebibliography}
\end{document}